\documentclass[prl,twocolumn]{revtex4-1}

\usepackage{graphicx}
\usepackage{amsmath}
\usepackage{epsfig}
\usepackage{dcolumn}  
\usepackage{bm}       
\usepackage{color}
\usepackage{epstopdf}

\newcommand{\sgn}{\mathop{\mathrm{sgn}}}

\begin{document}

\title{Transport, atom blockade and output coupling in a
  Tonks-Girardeau gas}

\author{L.~Rutherford$^1$, J.~Goold$^{2,3}$, Th.~Busch$^3$ and
  J.~F.~McCann$^1$}

\affiliation{$^1$ Centre for Theoretical Atomic, Molecular and Optical
  Physics, Queen's University Belfast, Belfast BT7 1NN, UK  \\ 
  $^2$ Clarendon Laboratory, University of Oxford, United Kingdom \\ 
  $^3$ Physics Department, University College Cork, Cork, Ireland}

\date{\today}

\begin{abstract}
  Recent landmark experiments have demonstrated how quantum mechanical
  impurities can be created within strongly correlated quantum gases
  and used to probe the coherence properties of these systems
  \cite{Palzer:09}.  Here we present a phenomenological model to
  simulate such an output coupler for a Tonks-Girardeau gas that shows
   qualitative agreement with the experimental results
  for atom transport and output coupling. Our model allows us to
  explore non-equilibrium transport phenomena in ultra-cold quantum
  gases and leads us to predict a regime of atom blockade, where the
  impurity component becomes localised in the parent cloud despite the
  presence of gravity. We show that this provides a stable
  mixed-species quantum gas in the strongly correlated limit.
\end{abstract}

\maketitle 
Ultra-cold gases provide an extremely versatile resource of quantum
matter. The possibility of applying external electromagnetic
potentials means that one can create macroscopic harmonic traps and
microscopic periodic structures that are sensitive to the quantum
states of the atomic species \cite{Bloch:08}. One recent breakthrough
has been the realisation that quasi one-dimensional quantum degenerate
gases can be created by using strong transverse trapping potentials
\cite{Moritz:03}. Such settings allow the radial (transverse) degrees
of freedom to be frozen out and in addition to tune the effective
one-dimensional interaction strength \cite{Olshanii:98}. The resulting
quantum many-body system of interacting bosons can be described by the
Lieb-Liniger model, which possesses exact solutions \cite{Lieb:63}.
In the limit of strong repulsive interactions these can be simplified
due to the existence of a mapping theorem to an equivalent gas of
non-interacting fermions \cite{Girardeau:60}. This so-called
Tonks-Girardeau (TG) limit was recently experimentally reached
independently in two separate laboratories \cite{Parades:04,Weiss:04}
and the initial experiments were followed by several spectacular
experimental studies.  These included the exploration of the
relationship between integrability and thermalization
\cite{Kinoshita:06} and also the creation and detection of meta-stable
excited states \cite{Haller:09}.

Very recently Palzer {\it et al.}~\cite{Palzer:09} demonstrated an
output coupler for an optically-trapped quantum gas in the TG
regime. An array of one-dimensional clouds was created in a
two-dimensional optical lattice and then probed using a
radio-frequency pulse to locally populate an untrapped hyperfine
level. The transport properties of this untrapped 'impurity' as it
fell under gravity, passing through the parent cloud, were
subsequently observed. This experiment constitutes a genuine open
quantum system and also one of the first experiments that explores the
important topic of quantum transport in a clean ultra-cold environment
(see also \cite{Fertig:05}). Here we propose a theoretical framework
that can describe the effects experimentally observed and predict the
existence of a regime of atom blockade, in which the interaction
between two components is strong enough to trap and localise the
impurity state. This phenomenon of self-trapping of neutral impurity
atoms in quantum degenerate gases has received a significant amount of
theoretical attention in recent times \cite{Cucchietti:06, Kalas:06,
  Bruderer:08} and our description paves the way for an immediate
experimental realization of the effect in this novel one-dimensional
configuration. Moreover, our work complements and extends recent
theoretical interest in the area of general impurity embedding in the
Tonks-Girardeau regime \cite{Goold:08, Minguizzi:09, Goold:10}.

A low-density gas of $N$ identical bosons trapped in a quasi
one-dimensional waveguide can be described by the Hamiltonian
\begin{eqnarray}
 \label{eq:TG_ham}
 \mathcal{H}=\sum_{i=1}^{N}
      \left[\frac{-\hbar^2}{2m}\frac{\partial^2}{\partial z_i^2}
      +V(z_i)\right]
      +\kappa \sum_{i<j}\delta(|z_i-z_j|),
\end{eqnarray}
where $m$ is the mass of the particles, $z$ the axial coordinate, and
$V (z_i) = \textstyle{1 \over 2} m\omega_{\parallel}^2 z_i^2$, is the
axial trapping potential, with $\omega_{\parallel}$ the corresponding
axial angular frequency and $a_{\parallel}$ is the corresponding trap
length.  The strength of the atom-atom contact interactions is given
by $\kappa$, the one-dimensional coupling constant, which can be
derived by a renormalisation procedure from the three-dimensional
scattering process as $\kappa =\frac{4\hbar^2 a_{3D}}{ma_\perp}
\left(a_\perp-Ca_{3D} \right)^{-1}$ \cite{Olshanii:98}. Here $a_\perp
= \sqrt{\hbar / m \omega_\perp}$ is the radial trap width and
$\omega_\perp$ the radial trapping frequency. The standard $s$-wave
scattering length is denoted by $a_{\rm 3D}$ and $C \approx
1.4603$. For the $^{87}$Rb isotope we have $a_{\rm 3D} \approx 5.3
\times 10^{-9}$m for both hyperfine states used in \cite{Palzer:09},
and $m \approx 1.44 \times 10^{-25}$kg.  For finite $\kappa$ the
Hamiltonian describes an inhomogeneous Lieb-Liniger gas and we
characterise the strongly interacing regime by demanding that the
contact interaction dominates the kinetic energy $\kappa \gg \hbar^2
n_{1D}/m$, where $n_{1D}$ is the mean linear density of the atoms. In
this limit the many body problem admits a unique and particularly
elegant solution, as it allows one to replace the contact interaction
in eq.~(\ref{eq:TG_ham}) with the nodal boundary condition
$\Psi_B(z_1,z_2,\dots,z_N,t)=0$ if $|z_i-z_j|=0$, for $i\neq j$ and
$1\leq i\leq\ j\leq N$. Such a constraint can be enforced a priori by
Slater determinant factorisation, $\Psi_F(z_1,z_2,\dots,z_N,t)
=\frac{1}{\sqrt N!}\det_{(n,j)=(0,1)}^{(N-1,N)}\psi_n(z_j,t)$, where
the $\psi_n$ are the single-particle eigenstates of the {\em
  non-interacting} system. We have adopted the convention of labeling
the first $N$ single-particle eigenfunctions with the index
$n=0,1,2,\dots N-1$. This, however, leads to a fermionic rather than
bosonic exchange symmetry, which is corrected by proper
symmetrisation, $A=\prod_{1\leq i < j\leq N} \sgn(z_i-z_j)$ to give
$\Psi_B=A \Psi_F$ \cite{Girardeau:60}. From this exact solution, the
time-dependent single-particle density is given by \cite{Wright:00,
  Yukalov:05}
\begin{eqnarray}
  \label{eq:spd}
  \rho(z,t)&=&N\int_{-\infty}^{+\infty}
              |\Psi_B (z,z_2,\dots,z_N;t)|^2 dz_2\dots dz_N
             \nonumber\\
           &=&\sum_{n=0}^{N-1}|\psi_{n}(z,t)|^2\;.
\end{eqnarray}
In the experiment by Palzer {\it et al.}~\cite{Palzer:09} an array of
one-dimensional clouds was created with the axis vertically aligned. A
localized excitation of a small section (near the center) of the cloud
to an untrapped hyperfine state was introduced.  While these untrapped
atoms fall under gravity, they remain strongly confined in the
transverse direction created by the optical fields. The dynamics of
the system can therefore be split into the description of its two
components: a trapped background gas of initially $N_\text{b}$ atoms,
and an impurity wave-packet of $N_\text{i}$ atoms, with atom densities
$\rho_\text{b}(z,t)$ and $\rho_\text{i}(z,t)$, respectively.

Girardeau originally used his famous mapping theorem to find the
unique solution to the single component many-body problem.  The
solution of the full time-dependent two-component many-body problem is
challenging and only a small number of exact solutions exist for time
independent mixtures \cite{girardeau:07}. Here we suggest that a
two-component system with transport can be treated using a
phenomenological approach based on the time evolution of non-linearly
coupled single particle states. The density distributions are then
built from these states by assuming single component Tonks-Girardeau
gases for both the impurity and parent components (see
eq.~\eqref{eq:spd}).

At time $t=0$, all atoms in the system are assumed to be in the
trapped TG gas state, the density of which can be calculated without
approximation using Eq.~\eqref{eq:spd}.  We then simulate the
experiment through the application of a short, broadband pulse,
$f(z,t)$, with a Gaussian spatial intensity of {\small FWHM}
$\sigma=2.3\mu$m, for a duration of $t_{\rm pulse}=200\mu$s. The
intensity of this pulse, $\gamma$, is chosen such that the population
transfer corresponds to the experimental situation in which, on
average, three atoms were coupled out \cite{Palzer:09}. The integrated
densities (norms) of the two components vary in time and according to
a set of coupled non-linear Schr\"odinger equations of the form
\begin{eqnarray}
 i\hbar{\partial \psi_n \over \partial t} &=&\left[ -\frac{\hbar^2}{2m}{\partial^2 \over \partial z^2}
  +\textstyle{1 \over 2} m\omega_{\parallel}^2 z^2 + \kappa\rho_\text{i} \right] \psi_n
  +\gamma f(z,t)\phi_n,\ \ \nonumber \\
 i\hbar{ \partial \phi_n \over \partial t} &=&\left[ -\frac{\hbar^2}{2m}{\partial^2 \over \partial z^2} 
  + mgz + \kappa\rho_\text{b} \right]\phi_n+\gamma f(z,t)\psi_n\;. 
 \label{eq:CoupledEqs}
\end{eqnarray}
The densities of the impurity and background components are then built
according to $\rho_{b}(z,t)=\sum_{n=0}^{N-1}|\psi_{n}(z,t)|^2$ and
$\rho_{i}(z,t)=\sum_{n=0}^{N-1}|\phi_{n}(z,t)|^2$ respectivley.  Here
$g$ is the gravitational acceleration and we have assumed that the
inter-component scattering is in mean-field regime. This can be
justified by realising that the impenetrability of the non-alike atoms
in the Tonks limit is overcome by the gravitational pull on the
impurity, which leads to an increase in the scattering parameter
$k|a_{1D}|$, where $k$ is the linear atomic momentum and $a_{1D}$ is the atomic scattering length.
In this regime the collisions between particles in different states
can no longer be treated as collisions between impenetrable
particles. The strength of this scattering, $\kappa$, is calculated
from the data given in \cite{Palzer:09} to be $\kappa \sim 1.35\times
10^{-36}$ Jm.  In the following we will show results from simulating
these coupled equations for a gas of initially $N_\text{b}=50$ atoms,
from which the outcoupling process transfers $N_\text{i}\approx 3$
into the impurity. All parameters are chosen such that a direct,
qualitative comparison to the experiment is possible \cite{Palzer:09}.

\begin{figure}[tb]
   \includegraphics[width=\linewidth]{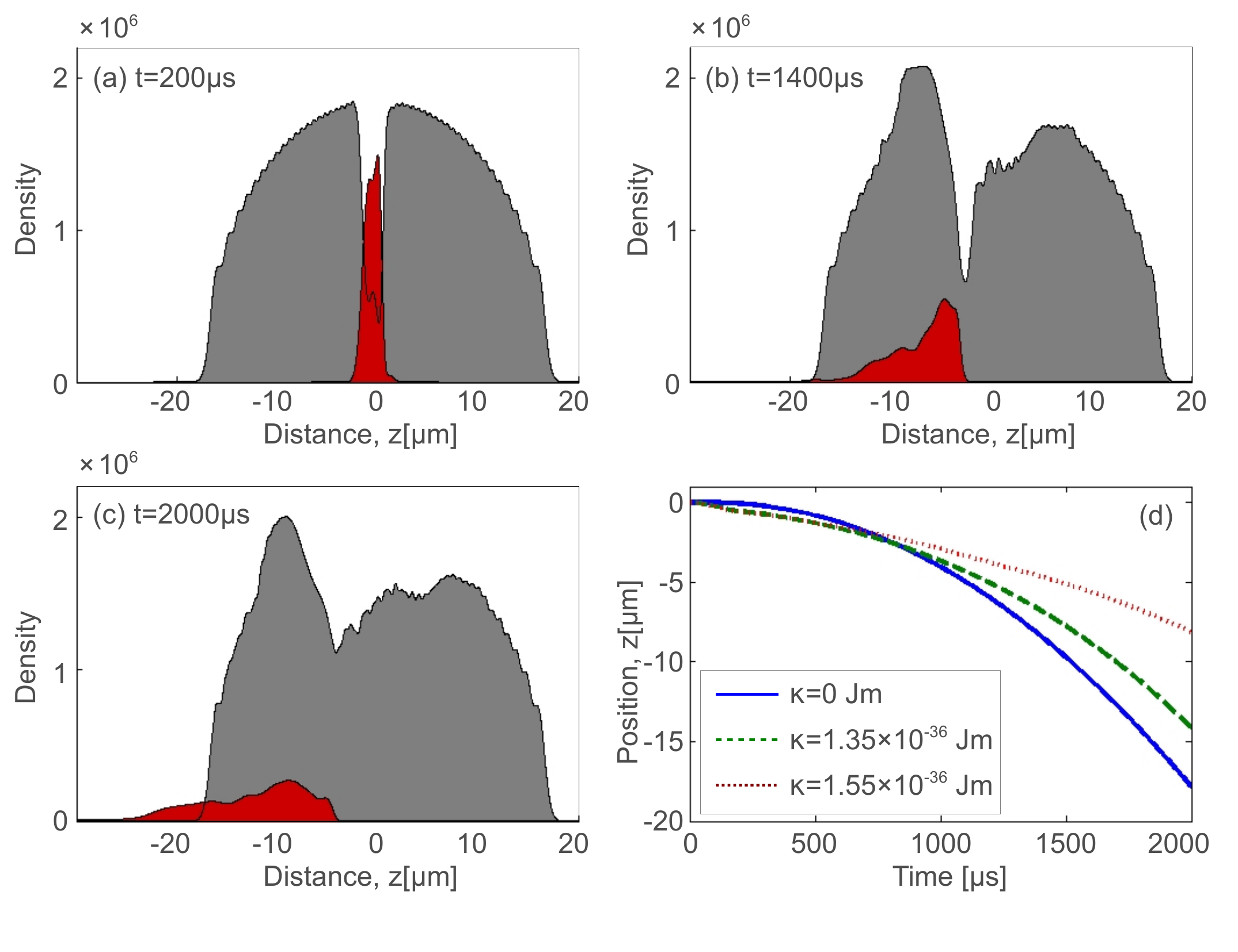}
   \caption{Density profiles of the two-component Tonks gas system for
     different times $(a)$ $t=200\,\mu$s, $(b)$ $t=1400\,\mu$s and
     $(c)$ $t=2000\,\mu$s for $N_b=50$ $^{87}$Rb atoms and
     $\omega_\parallel=2\pi\times39$ Hz. The grey (lighter) curves
     show the trapped component, while the red (darker) curves show
     the impurity wave-packet. The width of the outcoupling pulse is
     $\sigma=2.3\,\mu$m and the inter-component interaction strength
     is $\kappa=1.35\times10^{-36}$Jm.  Subfigure (d) shows the
     center-of-mass trajectory of the impurity atoms for different
     interaction strengths $\kappa$.}
  \label{fig:Densities}
\end{figure}

The densities of the components after the outcoupling process and
during the subsequent dynamics are shown in
Fig.~\ref{fig:Densities}. Initially, after the coupling pulse is
switched off ($t=200\mu$s), the impurity component is localised at the
origin ($z=0\mu$m) and subsequently disperses and displaces as it is
dragged through the trapped component by gravity ($t=1400\mu$s and
$t=2000\mu$s). The response of the background cloud on these
timescales is mainly determined by the interaction with the falling
impurity and our results agree qualitatively well with
the experimental findings of Palzer {\it et
  al.}~\cite{Palzer:09}. This provides support for our specific
phenomenological model, as other approaches, for example assuming
coherence within the impurity, led to a dynamics that agreed very
poorly.

The transport process was studied by analysis of the impurity's
center-of-mass motion, which is shown in Fig.~\ref{fig:Densities}(d)
for three different values of the interaction strength,
$\kappa$. It can be seen that a finite interaction
  produces noticeable deviations from parabolic flight. Firstly, the
center-of-mass is expelled downwards, on a very short time scale. This
can be understood by realising that the resonance position for the
coupling is moved towards negative $z$-values for increasing magnitude
of the non-linear terms in Eq.~\eqref{eq:CoupledEqs} (see also
discussion below). After this the fall of the outcoupled atoms is
slowed by the trapped cloud as compared to the free-fall case. We find
that for $\kappa=1.35\times10^{-36}$~Jm eventually all atoms exit the
overlap region with the trapped component, however the significant
slowdown for $\kappa=1.55\times10^{-36}$~Jm points towards the
interesting possibility of blockade within the gas.

\begin{figure}[tb]
 \includegraphics[width=\linewidth]{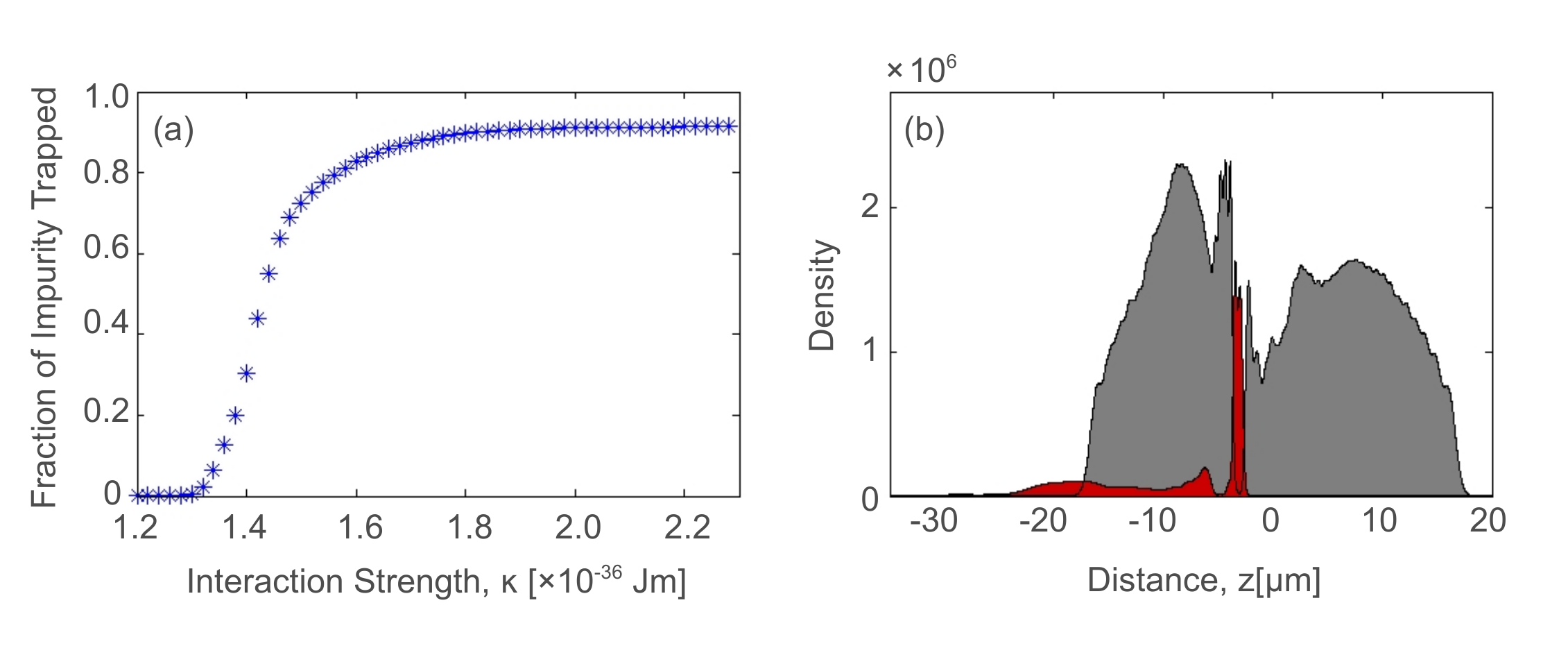}
 \caption{ (a) Atom blockade: shown is the fraction of out-coupled
   atoms remaining within the confines of the trapped gas after
   $t=3000\,\mu$s. (b) Density profile taken at $t=2000\,\mu$s for a
   gas in the intermediate regime at $\kappa=1.44\times10^{-36}$ Jm.}
  \label{fig:SelfTrapping}
\end{figure}

Being able to trap the impurity and leave it embedded in the parent
cloud is an exciting prospect for a novel mixed quantum gas. We
therefore calculate the fraction of impurity atoms that might be
retained within the confines of the trapped cloud ($[-20,20]\mu$m)
after $3000\,\mu$s. In each case approximately three atoms are present
in the impurity wave-packet and Fig.~\ref{fig:SelfTrapping}(a) shows
how the fraction remaining within the cloud region varies as a
function of $\kappa$. Three regimes can be identified: (i) at
interaction strengths below $\kappa\approx 1.35\times10^{-36}$~Jm the
entire impurity wave-packet is able to pass through the trapped
component and exit the cloud; (ii) in the region between
$\kappa\approx 1.35\times10^{-36}$~Jm and $\kappa\approx
1.5\times10^{-36}$~Jm the impurity wave-packet splits into two
components of which one leaves the cloud and the other remains trapped
in the center; and (iii) above $\kappa\approx 1.5\times10^{-36}$~Jm
maximum self-trapping is achieved, with $>90\%$ of the atomic density
blockaded inside the trapped gas for long timescales. A typical
density plot for an interaction strength in the intermediate regime
($\kappa=1.44\times10^{-36}$~Jm) after $2000 \mu$s is shown in
Fig.~\ref{fig:SelfTrapping}(b). Approximately half of the out-coupled
density is clearly visible to be localised in the center of the
background gas, while the other half has been significantly
accelerated by gravity. This part will continue to travel out of the
cloud.

\begin{figure}[tb]
  \includegraphics[width=\linewidth]{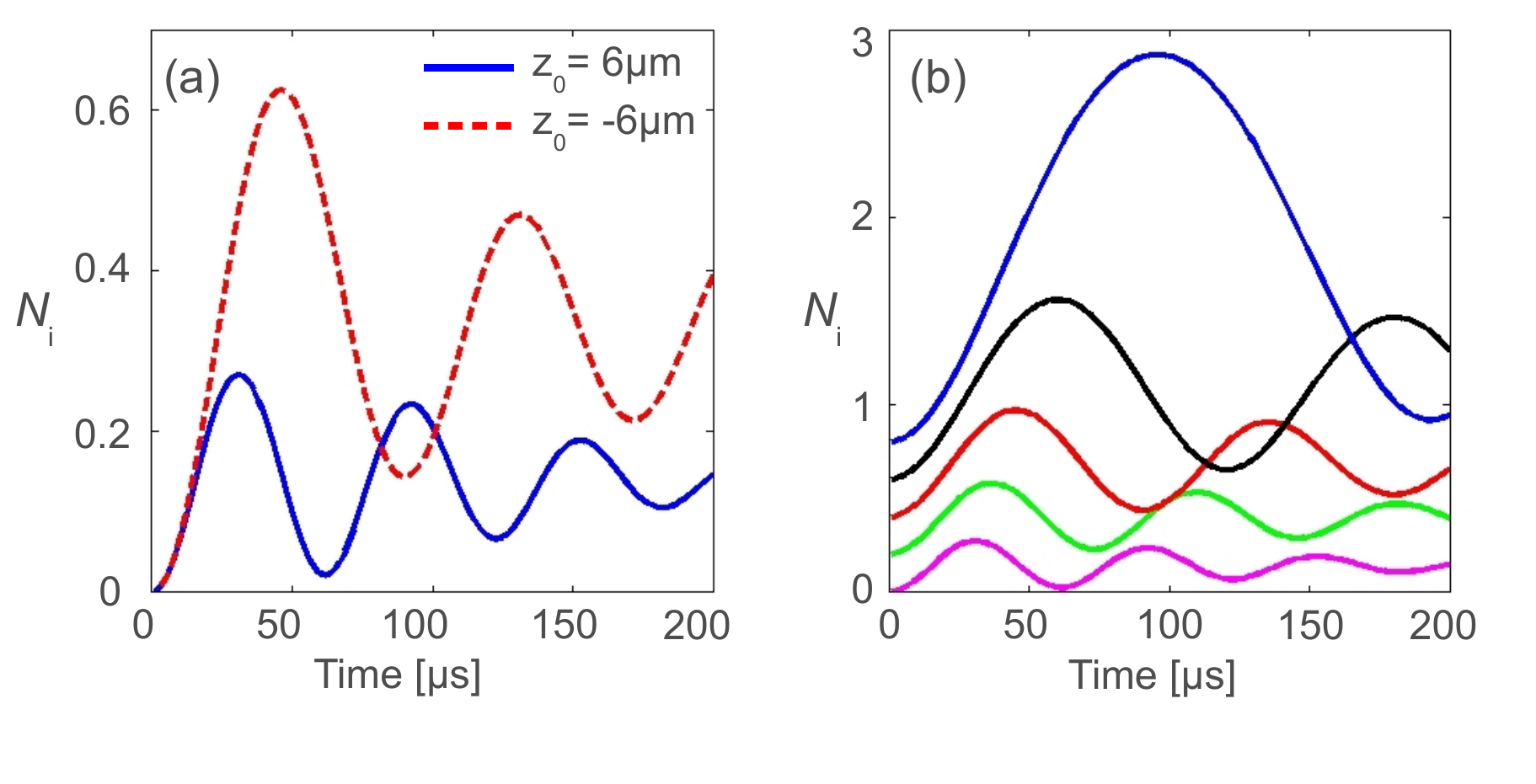}
  \caption{Output yield for a pulse of duration $t_{\rm pulse} =
    200\mu$s. Panel (a) shows the asymmetry present when the focus of
    the pulse is offset by a distance $z_0=\pm 6\mu$m from the center
    of the cloud.  Panel (b) shows how the frequency of the Rabi
    oscillations (at $z_0=6\mu$m) are affected for different strength
    of gravity, $g_\text{eff}/g=0.2,0.4,0.6,0.8,1$ corresponding to
    the curves from top to bottom. The curves are vertically offset
    for clarity by $\Delta N=0.2$. Note that the
      justification for the non-linear term in
      eq.~\eqref{eq:CoupledEqs} requires a finite gravitational
      strength. }
  \label{fig:OffCentreCoupling}
\end{figure}

Let us in the following consider the effects due to the asymmetry
induced by the gravitational potential.  For this we look at the
situation in which the centre of the outcoupling pulse is located away
from the origin of the harmonic potential.
Fig.~\ref{fig:OffCentreCoupling}(a) shows the results for the number
of outcoupled atoms when the pulse is focused around $z_0=\pm
6\mu$m. One immediately notes that the outcoupling yield at both
positions is dramatically reduced in comparison with coupling from the
center.  While this can be partly explained by the lower density of
the TG gas at $z=\pm 6\mu$m, it alone is not sufficient for the
dramatic reduction. In fact, the low efficiency is due to the
increasingly off-resonant nature of the outcoupling process for larger
distances from the centre, which stems from the presence of the strong
gravitational potential in the outcoupled channel. This is also
consistent with the increased value of the Rabi frequency as can be
seen from Fig.~\ref{fig:OffCentreCoupling}(a) for $z_0=6\mu$m. In
fact, carrying out simulations for a system in which the strength of
gravity can be decreased (corresponding to the optical lattice being
rotated from vertical to horizontal), shows in increase in Rabi
frequencies with decreasing detuning between the channels (see
Fig.~\ref{fig:OffCentreCoupling}(b)). Note that while during the time
of the outcoupling process the gravitational pull has only a small
effect on the position of the outcoupled atoms, the dephasing of the
output can already be seen in the damping of the Rabi oscillations.

\begin{figure}[tb]
  \includegraphics[width=\linewidth]{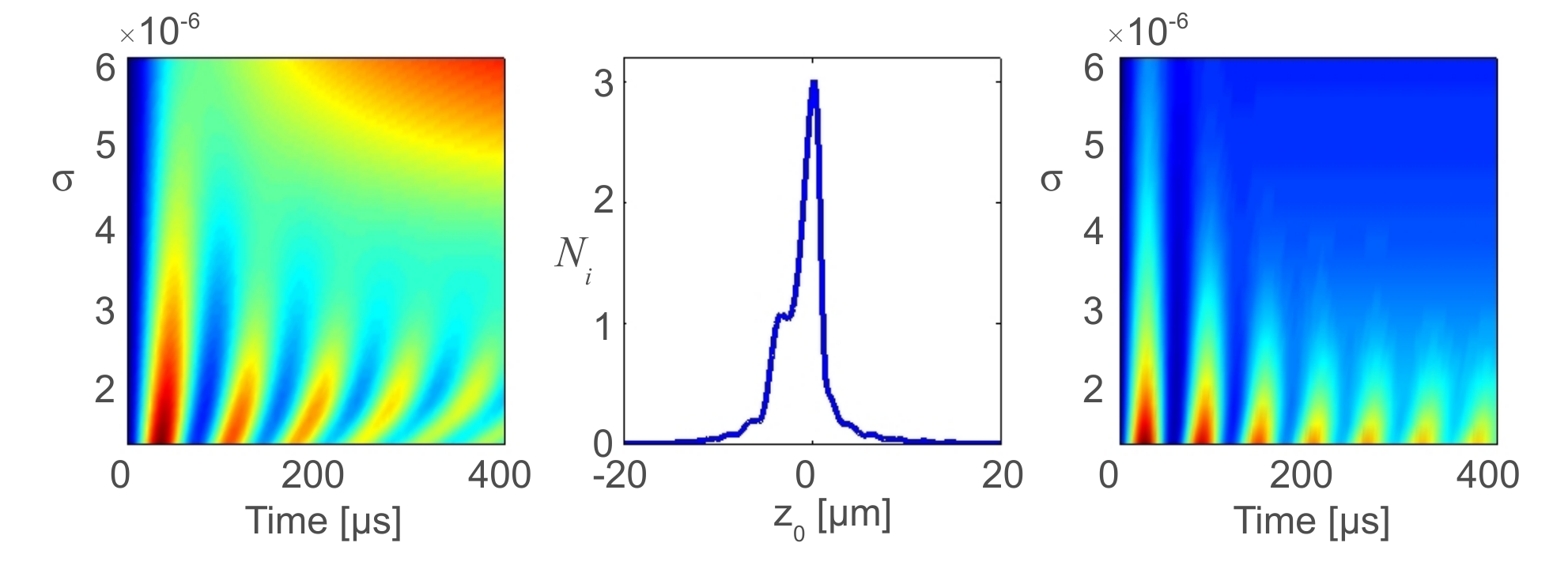}
  \caption{The effect of pulse width $\sigma$ on the atom output yield
    from a TG gas below and above the centre of the trap. Left: atom
    yield for a pulse focused at $z_0=-6\mu$m.  Center: Output yield
    (at the end of the pulse) as a function of focal point for
    $\sigma=2.3\mu$m. Right: yield for focus at $z_0= 6\mu$m.
    Well-defined Rabi oscillations are clearly visible for small
    $\sigma$ for for both situations, above (right) and below (left)
    the trap centre.}
\label{fig:Pulsewidth}  
\end{figure}

To understand the yield variations observed in
Fig.~\ref{fig:OffCentreCoupling}(a) let us in the following
investigate the reaction of the system to a change in the width,
$\sigma$, of the outcoupling pulse.  In the middle panel of
Fig.~\ref{fig:Pulsewidth} we show the output yield at the end of the
coupling pulse ($t=200\mu$s) as a function of the position of the
focal point for a pulse of the experimental {\small FWHM} of
$\sigma=2.3\mu$m.  Focusing on or close to the center of the trap
ensures the expected large efficiencies, and the strong asymmetry
observed in Fig.~\ref{fig:OffCentreCoupling}(a) becomes visible for
increasing values of $|z_0|$. The visible fine-structure is due to the
existence of Rabi oscillations. The strong fall-off away from the trap
centre shows that the effect of gravitational detuning is important
over the scale of a few microns and therefore can be important over
the spatial profile of the outcoupling pulse. It will, in particular,
influence the coherence of the coupling process and we show in the
left and right panels of Fig.~\ref{fig:Pulsewidth} that an increase of
the pulse width, $\sigma$, leads to a decay of the coherent,
high-contrast Rabi oscillations. The fact that for outcoupling pulses
of the same duration the number of oscillations visible for small
values of $\sigma$ is different for coupling above and below the cloud
centre can be understood by remembering that the resonance point is
below the the trap centre. Therefore the detuning at positive $z_0$ is
larger than at negative $z_0$, leading to higher frequency
oscillations.

As the spot size increases the detuning gradient gives rise to
different phase components in the outcoupled pulse, which leads to the
vanishing of the Rabi oscillations. This is also the reason for the
displacement of the oscillations with increasing spot size, as broader
spots are able to sample areas closer to the resonance (which for our
parameters lies at $z_R=-1.9\mu$m) with lower Rabi frequency. As the
spot size increases further a pulse at $z_0=-6\mu$m significantly
overlaps with the resonance point and its output becomes dominated by
a single (coherent) component from the high-density central region of
the trap. This leads to the increase observed in the upper right hand
side corner of the left panel of Fig.~\ref{fig:Pulsewidth}. While the
spot size for the outcoupling pulse cannot be made arbitrarily small,
these effects could be experimentally observed by weakening the
longitudinal trapping, for example.

In conclusion, we have presented a model for transport in strongly
correlated quantum gases that is able to reproduce key features of
recent experiments. Furthermore, our model predicts several features
which could be explored further experimentally. We have brought
strong evidences that even in the presence of gravity a strong interaction between an
out-coupled and a residing component in a Tonks gas can lead to a self
localization of the out-coupled component. Although atom blockade was
not observed in the experiment of Palzer {\it et al.}, the regime
could be reached by tuning a Feshbach resonance between the hyperfine
states. In addition our work has shed light on interesting aspects of
the out coupling process which are unique to the configuration under
consideration. Using our model to investigate other condensed matter
phenomena in these correlated systems, such as spin-charge separation
and analogues of Cherenkov radiation, are subjects of ongoing
research.

\begin{acknowledgments}
  LR is grateful to the Department for Employment and Learning
  (Northern Ireland) for a PhD studentship. The work was supported by
  Science Foundation Ireland under project number 05/IN/I852 and
  05/IN/I852 NS. JG would like to acknowledge funding from an IRCSET
  Marie Curie International Mobility fellowship.  JG and TB would like
  to thank A.~del Campo for interesting discussions.
\end{acknowledgments}


\begin{thebibliography}{99}

\bibitem{Palzer:09} S.~Palzer, C.~Zipkes, C.~Sias and M.~K\"ohl,
  Phys.~Rev.~Lett.~{\bf 103}, 150601 (2009).

\bibitem{Bloch:08} I.~Bloch, J.~Dalibard and W.~Zwerger,
  Rev.~Mod.~Phys.~{\bf 80}, 885 (2008).

\bibitem{Moritz:03} H.~Moritz, T.~St\"oferle, M.~K\"ohl and
  T.~Esslinger, Phys.~Rev~Lett.~{\bf 91}, 250402 (2003).

\bibitem{Olshanii:98} M.~Olshanii, Phys.~Rev.~Lett.~\textbf{81}, 938
  (1998).
  
\bibitem{Lieb:63} E.~Lieb and W.~Liniger, Phys.~Rev.~\textbf{130},
  1605 (1963).

\bibitem{Girardeau:60} M.~Girardeau, J.~Math.~Phys.~\textbf{1}, 516
  (1960).

\bibitem{Parades:04} B.~Paredes, A.~Widera, V.~Murg, O.~Mandel,
  S.~F\"olling, I.~Cirac, G.V.~Shlyapnikov, T.W.~H\"ansch and
  I.~Bloch, Nature \textbf{429}, 277 (2004).

\bibitem{Weiss:04} T.~Kinoshita, T.~Wenger and D.S.~Weiss, Science
  \textbf{305}, 1125 (2004).
  
\bibitem{Kinoshita:06} T.~Kinoshita, T.~Wenger and D.S.~Weiss, Nature
  \textbf{440}, 900 (2006).
  
\bibitem{Haller:09} E.~Haller, M.~Gustavsson, M.~J.~Mark, J.G.~Danzl,
  R.~Hart, G.~Pupillo and H.C.~N\"agerl, Science \textbf{325}, 1224
  (2009).

\bibitem{Fertig:05} C.D.~Fertig, K.M.~O Hara, J.H.~Huckans,
  S.L.~Rolston, W.D.~Phillips and J.V.~Porto, Phys.~Rev.~Lett.~{\bf
    94}, 120403 (2005).

\bibitem{Cucchietti:06} F.M.~Cucchietti and E.~Timmermans,
  Phys.~Rev.~Lett.~{\bf 96}, 210401 (2006).

\bibitem{Kalas:06} Ryan M.~Kalas and D.~Blume, Phys.~Rev.~A {\bf 73},
  043608 (2006).
 
\bibitem{Bruderer:08} M.~Bruderer, W.~Bao and D.~Jaksch,
  Europhys.~Lett.~{\bf 82}, 30004 (2008).

\bibitem{Goold:08} J.~Goold and Th.~Busch, Phys.~Rev.~A {\bf 77},
  063601 (2008).

\bibitem{Minguizzi:09} M.~Girardeau and A.~Minguizzi,
  Phys.~Rev.~A {\bf 79}, 033610 (2009).

\bibitem{Goold:10} J.~Goold, H.~Doerk, Z.~Idziaszek, T.~Calarco and
  Th.~Busch, Phys.~Rev.~A {\bf 81}, 041601 (2010).

\bibitem{Yukalov:05} V.I.~Yukalov and M.D.~Girardeau, Laser
  Phys.~Lett.~{\bf 2}, 375 (2005).

\bibitem{girardeau:07} M.~Girardeau and A.~Minguizzi, Phys.~Rev~Lett. {\bf 99}, 230402 (2007).

\bibitem{Wright:00} M.D.~Girardeau and E.M.~Wright,
  Phys.~Rev.~Lett.~{\bf 84}, 5239 (2000).

\bibitem{Fleck} M.~Feit, J.A.~Fleck Jr. and A.~Steiger,
  J.~Comput.~Phys.~{\bf 47}, 412 (1982).

\bibitem{Girardeau:09} M.D.~Girardeau and A.~Minguzzi, Phys.~Rev.~A
  {\bf 79}, 033610 (2009).

\end{thebibliography}
\end{document}